\documentstyle[twocolumn,aps,prl,overcite,epsfig]{revtex}

\begin{document}

\def\simg{\mathrel{\hbox{\rlap{\lower.55ex \hbox {$\sim$}}
                   \kern-.3em \raise.4ex \hbox{$>$}}}}
\def\siml{\mathrel{\hbox{\rlap{\lower.55ex \hbox {$\sim$}}
                   \kern-.3em \raise.4ex \hbox{$<$}}}}
\def\bitm{\bibitem}
\def\beq{\begin{equation}}
\def\enq{\end{equation}}
\def\bea{\begin{eqnarray}}
\def\ena{\end{eqnarray}}
\def\nonum{\nonumber}
\def\bec{\begin{center}}
\def\enc{\end{center}}
\def\etal{{\it et al.}}
\def\to{\rightarrow}
\def\Mesz{M\'esz\'aros~}
\def\barnu{\bar\nu}
\def\barnue{\bar \nu_e}
\def\barnum{\bar \nu_\mu}
\def\epm{\hbox{e}^\pm}
\def\nue{\nu_e}
\def\num{\nu_\mu}
\def\E53{E_{53}}
\def\L52{L_{52}}
\def\tw1{t_{w1}}
\def\ro7{r_{o7}}
\def\Gpf{\Gamma_{pf}}
\def\Gnf{\Gamma_{nf}}
\def\msun{M_\odot}
\def\eps{\epsilon}
\def\Tho{\Theta_o}
\def\Th{\Theta}
\def\st{\sigma_T}
\def\ets{\eta_\ast}
\def\etsg{\eta_{\ast\gamma}}
\def\etspi{\eta_{\pi}}

\title{5-10 GeV Neutrinos from Gamma-Ray Burst Fireballs}

\author{John N. Bahcall$^1$ \& Peter \Mesz$^{1,2}$ \\
$^1$ Institute for Advanced Study \\
$^2$ Pennsylvania State University }

\maketitle


\begin{abstract}
A gamma-ray burst fireball is likely to contain an admixture of neutrons, in 
addition to protons, in essentially all progenitor scenarios.
Inelastic collisions between differentially streaming protons and neutrons 
in the fireball produce $\num(\barnum)$ of $\sim$10 GeV as well as 
$\nue(\barnue)$ of $\sim 5$ GeV, which could produce $\sim 7$ events/year 
in km$^3$ detectors, if the neutron abundance is comparable to that of protons.
Photons of $\sim 10$ GeV from $\pi^0$ decay and $\sim 100$ MeV $\barnue$ from neutron 
decay are also produced, but will be difficult to detect. Photons with energies
$\siml 1$ MeV from shocks following neutron decay produce a characteristic signal
which may be distinguishable from the proton-related MeV photons.
\end{abstract}

\pacs{PACS numbers: 96.40.Tv,98.70.Rz,98.70.Sa,14.60.Pq }

\section{Introduction}

Gamma-ray burst (GRB) sources are distributed throughout the universe and their 
energy output is measured to be a substantial fraction of a solar rest mass 
equivalent \cite{fm95}.  A variety of observations support the interpretation that these 
events are caused by cataclysmic stellar collapse or compact mergers, producing
a fireball with bulk expansion Lorentz factor $\Gamma \sim 10^2-10^3$. 
In the standard GRB model a fireball made up of $\gamma,\epm$ 
and magnetic fields
with an admixture of baryons is produced by the release of a large amount of energy 
$E\simg 10^{53}$ ergs in a region $r_o \sim 10^7 \ro7$ cm (e.g. \cite{pm99}).
The observations indicate that typical fireballs 
are characterized by a luminosity $L\sim 10^{52}\L52$ erg s$^{-1}$ and durations $t_w = 
10 t_{w1}$ s in the observer frame, with a large spread in both quantities. The outflow 
is controlled by the value of the dimensionless entropy  $\eta=(L/{\dot M} c^2)$ 
injected at $r_o$. Previous discussions of fireball models have generally focused on the 
charged particle components, since they determine directly the photon signal. 
However, consideration of a neutron component introduces qualitatively new effects 
\cite{dkk99}.  In a $p,n$ fireball, for values of $\eta\simg 400$, the neutrons and 
protons acquire a relative drift velocity causing inelastic $n,p$ collisions and creating 
neutrinos. 

We investigate here the neutrino and photon signals from $n,p$ collisions following 
decoupling in GRB. The $\sim 10$ GeV neutrinos from this mechanism depend upon the presence 
of neutrons in the original explosion, but the neutrinos are created in simple physical
processes occurring in the later stages of the fireball.
On the other hand, the $10^5$ GeV neutrinos discussed in 
refs. \cite{wb97} require the acceleration in shock waves of ultra-high energy protons
interacting with photons. Thus the 10 GeV and the $10^5$ GeV neutrinos reflect very 
different astrophysical processes and uncertainties. Other processes, e.g. neutrinos from 
$p,p$ collisions\cite{px94} also require shocks but have lower efficiencies, while 10-30 
MeV neutrinos from the original explosion\cite{pk99} are much harder to detect due to the
lower cross sections.

We show (\S \ref{sec:nudet}) that the 10 GeV neutrinos could be 
detectable by future km$^3$ size detectors. The associated $\sim 10$ GeV $\gamma$-ray
fluences are compatible with current detection rates, and may be detectable with 
future space missions.  The dependence of these signals on the neutron/proton ratio $\xi$ 
provides a new tool to investigate the nature of the GRB progenitor systems. Moreover 
the predicted neutrino event rate depends on the asymptotic bulk Lorentz factor of the 
neutrons, which is linked to that of the protons.  The latter affects all the 
electromagnetic observables from the GRB fireball, including the photospheric and 
shock radii, as well as the particle acceleration and non-thermal photon production.

\section{Dynamics, n-p Decoupling and Pions}
\label{sec:decoup}

Above the fireball injection radius $r_o$ the outflow velocity increases through conversion 
of internal energy into kinetic energy, the bulk Lorentz factor $\Gamma$ varying as 
$\Gamma \sim T'_o/T' = r/r_o$, where $T'$ is the comoving temperature and $T'_o =
1.2 \L52^{1/4}\ro7^{-1/2}$ MeV is the initial temperature at $r_o$ (henceforth denoting
with primes quantities measured in the comoving frame).  
The flow may be considered spherical, which is a valid approximation also for a
collimated outflow of opening angle $\theta_j > \Gamma^{-1}$, for the conditions 
discussed here.
In a pure proton
outflow the linear growth of $\Gamma$ saturates when it reaches an asymptotic value 
$\Gamma_f \leq \eta\sim$ constant, the value $\eta$ being achieved when the fireball 
converts all its luminosity into expansion kinetic energy. For an $n,p$ fireball, 
beyond the injection radius $r_o$ the comoving temperature is low
and nuclear reactions are rare, so the $n/p$ ratio $\xi$ remains
constant.  Since the thermal velocities are non-relativistic,
decoupling of the $n$ and $p$ fluids is essential for high-energy
neutrino production.

\begin{figure}[htb]
\centering
\epsfig{figure=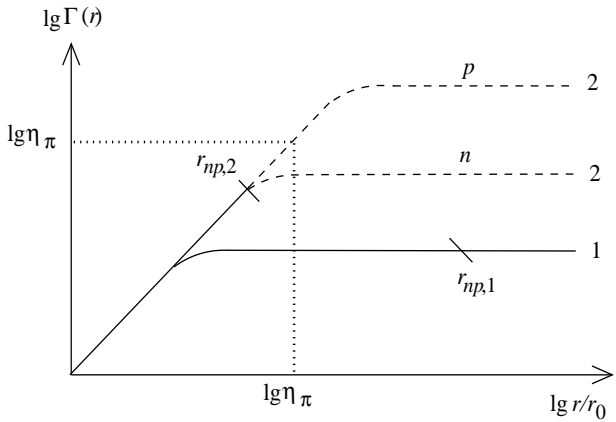, width=3.2in, height=2.2in}
\vspace*{.1in}
\caption{
Schematic behavior of the bulk Lorentz factor $\Gamma$ as a function of radius $r$ for 
various values of the dimensionless entropy $\eta$, the decoupling radius  $r_{np}$ being
indicated with a diagonal slash. Curve 1 is for $\eta < \etspi$, where the $n$ and $p$ 
achieve the same asymptotic $\Gnf\sim\Gpf\sim\eta$.  Curve 2 is for $\eta >\etspi$, and 
in this case $n,p$ decoupling occurs before protons have reached their asymptotic Lorentz 
factor, which is larger than that of neutrons. This leads to inelastic $n,p$ collisions, 
pion formation and neutrino emission at $r_{np,2}$. 
   \label{fig:1}
}
\end{figure}

At the base of the outflow the $n$ and $p$ components are coupled 
by nuclear elastic scattering. In terms of the CM relative energy $\eps_{rel}$ and the
relative velocity $v_{rel}$ between nucleons, $\sigma'_{el} v_{rel}' \sim \sigma_o c$.
The CM energy dependence $\sigma_{el} \propto \eps'^{-1/2} \propto v_{rel}'^{-1}$ is 
approximately valid between energies $\sim$ MeV and the pion production threshold $\sim 140$ 
MeV, and $\sigma_o \sim \sigma_\pi \sim 3\times 10^{-26}$ cm$^2$ is the pion formation cross 
section above threshold.  The $p$ and $n$ are cold in the comoving frame, and remain well 
coupled until the comoving $n,p$ scattering time $t'_{np} \sim (n'_p \sigma_o c)^{-1}$ 
becomes longer than the comoving expansion time $t'_{exp}\sim r/c\Gamma$. 
Denoting the comoving neutron density $n'_n=\xi n'_p$ with $\xi\siml 1$, mass conservation 
implies $n'_p=L/[(1+\xi)4\pi r^2 m_p c^3 \Gamma \eta ]$. The $n,p$ decoupling occurs in
the coasting or accelerating regimes depending on whether the dimensionless entropy 
$\eta$ is below or above the critical value
\bea
\etspi= & \left( { L \sigma_\pi / 4\pi m_p c^3 r_o (1+\xi)}\right)^{1/4}\qquad\qquad\nonum\\
 \simeq & 3.9\times 10^2 \L52^{1/4}\ro7^{-1/4} \left( {[1+\xi] / 2} \right)^{-1/4}~.
\label{eq:etspi}
\ena
Figure \ref{fig:1} shows the dependence of $\Gamma$ on radius for different $\eta$.
For low values, $\eta\siml\etspi$, the condition $t'_{np}\simg t'_{exp}$ is achieved 
at a radius $r_{np}/r_o =\etspi(\etspi/\eta)^3$, which is beyond the saturation radius 
$r_s/r_o\sim\eta$ at which both $n$ and $p$ start to coast with $\Gamma\sim \eta =$ constant. 
In this case, even after decoupling both $n$ and $p$ continue to coast together due to 
inertia, and their relative velocities never reach the threshold for inelastic collisions. 
%

For $\eta\simg\etspi$, on the other hand, the $n,p$ decoupling condition 
$t'_{np} \simg t'_{exp}$ occurs while the protons (and neutrons) are still accelerating 
as $\Gamma_p \simeq (r/r_o)$, at a radius
\beq
(r_{np}/ r_o) = \etspi (\eta/\etspi)^{-1/3}~,~~~\hbox{for}~~\eta\simg\etspi~.
\label{eq:rpi}
\enq
Beyond this decoupling radius the $p$ can still continue to accelerate with 
$\Gamma_p\propto r$ (as long as they remain coupled to the photons). However the neutrons are 
no longer accelerated, since they only interact with the protons, and they continue to coast 
with the value of $\Gamma\sim\Gnf\simeq$ constant achieved up to that point,
\bea
\Gnf= & (3/4) \etspi (\eta/\etspi)^{-1/3}\qquad (\hbox{for}~~~\eta\simg\etspi)\nonum\\
    \simeq & 3\times 10^2 
      \L52^{1/4}\ro7^{-1/4} \left( {[1+\xi ]/ 2} \right)^{-1/4}(\eta/\etspi)^{-1/3}~,
\label{eq:Gnf}
\ena
where the (3/4) factor comes from a numerical solution \cite{dkk99} of the coupling equations. 

When the $n,p$ decoupling condition $\eta\simg\etspi$ is satisfied, the relative $n,p$ drift 
velocity $v_{rel} \to c$ and the inelastic pion production threshold $\eps' >$ 140 MeV is 
reached. Since $\sigma_o
\sim\sigma_\pi$, the condition $t'_{np}\sim t'_{exp}$ implies that the optical depth to 
pion formation is of order unity. Thus, for $\eta\simg\etspi$ the radius $r_{np}\equiv r_\pi$
is not only a decoupling radius but also an effective ``pionospheric" radius.  

The lowest energy threshold processes at $r_\pi$ are
\goodbreak
\bea
p+n & \to  p+p+ \pi^- & \to \mu^-  + \barnum    \to e^-  + \barnue + \num +\barnum \nonum \\
    & \to  n+n+ \pi^+ & \to \mu^+  + \num       \to e^+  + \nue + \barnum +\num \nonum\\
    & \to  p+n+ \pi^0 & \to \gamma + \gamma     ~, 
\label{eq:pncol}
\ena
which occur in approximately equal ratios and with near unit total probability. 
The corresponding $p+p~(n+n) \to \pi^\pm, \pi^0$ 
processes do not involve a relative drift velocity (as do the $p+n$), and are thus
less probable. Processes leading to multiple baryons are also suppressed due to the 
higher threshold, and for simplicity we restrict ourselves to the above $p+n$ processes.

\section{10 GeV Neutrinos and $\gamma$-rays}
\label{sec:nudet}

The total number of neutrons carried by the fireball is
\bea
N_n=& \left({\xi\over 1+\xi}\right) {E\over \eta m_p c^2 } \qquad\qquad\qquad \nonum\\
\sim & 0.83\times 10^{53}\E53 \left({2\xi \over 1+\xi}\right)\left({400 \over \eta}\right)~,
\label{eq:Nn}
\ena
The comoving optical depth $\tau'\sim n'_p \sigma r/\Gamma\propto \sigma/(r\Gamma)$ has the 
same dependence for pion formation and photon scattering, but $\sigma_\pi \ll \st$ (Thomson 
cross section), so the pionosphere $r_\pi$ occurs below the $\gamma$-photosphere $r_\gamma$.
The $\gamma$-rays in equation (\ref{eq:pncol}) can only escape from a skin depth below
the $\gamma$-sphere in the essentially laminar flow 
with probability $P_\gamma\siml \tau_\pi 
(r_\gamma) \sim r_\pi/r_\gamma \sim (\sigma_\pi/\st)(1+\xi/7)^2 \sim 1/25$, for 
$\eta\simg\etspi$. Each $n$ leads to $\sim 1$
photon of CM energy $\eps'_\gamma\sim 70$ MeV and observer energy centered broadly 
around $\eps_\gamma\sim 70\Gnf/(1+z)\hbox{MeV}\sim 10$ GeV. Using a proper distance 
$D_p=2.8\times 10^{28} h_{65}^{-1} [1-1/\sqrt{1+z}]$ cm with a Hubble constant $h_{65}=
H_o/65\hbox{Km/s/Mpc}$, the number fluence at Earth is $N_\gamma\sim N_n P_\gamma/4\pi D_p^2 
\sim 10^{-5}$ cm$^{-2}$.  
This is below the sensitivity of the $\sim 200$ cm$^2$ area EGRET detector on the 
Compton Gamma Ray Observatory (e.g. \cite{cat97}), but for rare nearby bursts it may be
detectable by GLAST \cite{glast99}.

The neutrinos originate at the pionospheric radius $r_\pi\ll r_\gamma$ where 
$\tau_\pi\sim 1$. In this region the stable charged products and $\gamma$-rays from the 
reactions (\ref{eq:pncol}) remain in the fireball, and each $n$ leads on average to one 
$\nu$ and one $\barnu$. 
We list below the average neutrino energies for pions and muons decaying at rest. The
neutrinos from muon decay have a continuum spectrum. Also, the energies are Doppler
broadened by $v_{rel}/c\sim 0.5$.  
\bea
\label{eq:epsnucm}
\eps'_{\barnum}\simeq& 30\hbox{MeV}~~,~~\eps'_{\num}\simeq 30\hbox{MeV}
                                                \hbox{\quad from\quad }\pi^\pm \nonum\\
\eps'_{\nue}\simeq & 30 \hbox{MeV} ~~,~~\eps'_{\barnum}\simeq 50\hbox{MeV}
                                                 \hbox{\quad from\quad}\mu^+~,\\
\eps'_{\barnue}\simeq & 30\hbox{MeV}~~,~~ \eps'_{\num}\simeq 50\hbox{MeV} 
                                                 \hbox{\quad from\quad}\mu^- \nonum
\ena

The relevant cross section for detection averaged over $\nu$ and $\barnu$ is $\sigma_\nu 
\sim 0.5 \times 10^{-38} (\eps/\hbox{GeV})$ cm$^2$ at the observed energy $\eps$ 
\cite{gai90}.
The observer frame energy is $\eps= \eps' \alpha \Gnf /(1+z)$, where $\alpha\sim 1$ near
threshold. For the CM $\nu\barnu$ production energies of equation (\ref{eq:epsnucm}), the 
average $\nu+\barnu$ CM energy per neutron is $\eps'\simeq 100$ MeV. Taking $\alpha\simeq 
1$ the observer $\nu+\barnu$ energy per neutron is $\eps\simeq 0.1 \Gnf/(1+z)$ GeV, 
and the effective detection cross section per neutron is ${\overline \sigma_{\nu\barnu} } 
\sim 5\times 10^{-40} \Gnf (1+z)^{-1}$ cm$^2$.  Multiplying by a burst rate within a 
Hubble distance of ${\cal R}_b\sim 10^3 {\cal R}_{b3}$/year, for a 1 km$^3$ detector 
containing $N_t\sim 10^{39} N_{t39}$ target protons, the rate 
$R_{\nu\barnu}=(N_t /4\pi D_p^2 ){\cal R}_b N_n {\overline \sigma_{\nu\barnu}}$ is
\bea
R_{\nu\barnum}\sim  & 7 \E53 N_{t39}{\cal R}_{b3} 
   \left({2 \xi \over 1+\xi}\right) \left({\etspi \over \eta }\right)^{4/3}\nonum\\
   & \times h_{65}^2 \left( {2- \sqrt{2} \over 1+z - \sqrt{1+z}}\right)^2~\hbox{year}^{-1}~,
\label{eq:Rnubarnu}
\ena
events in the detector in coincidence with GRB electromagnetic flashes. 
The energies of the events are
\beq
\eps_{\num\barnum} \sim 10 \hbox{GeV}~,~
\eps_{\nue\barnue}\sim  5 \hbox{GeV}~,
\label{eq:epsnubarnu}
\enq
which scale $\propto \E53^{1/4}\tw1^{-1/4}\ro7^{-1/4}$ 
$(2 /[1+\xi])^{1/4} (2/[1+z])$ $(\etspi/\eta)^{1/3}$. 

Subsequent to decoupling and $n,p$ collisions, each neutron decay $n\rightarrow p+e^- 
+\barnue$ leads to an additional $\barnue$ of CM energy $\eps'_{\barnue,d}\sim 0.8$ MeV, 
which boosted in the observer frame by $\Gnf/(1+z)$ is $\siml 120 MeV$. 
The cross section is $\sigma_{\barnue}\sim 2\times 10^{-40}$ cm$^2$ and the
expected rate in a km$^3$ detector is less than one event per year.

\section{MeV $\gamma$-Rays} 
\label{sec:phot}

The non-thermal MeV $\gamma$-rays are thought to be produced in collisionless 
shocks\cite{rm94}, which occur at a radius $r_{sh} > r_\gamma > r_\pi$, after the bulk 
Lorentz factor has saturated to its asymptotic value. For an $n,p$ outflow, shocks can 
occur in the original $p$, as well as in the $n$ component after the latter have decayed,
and this can influence the external shock light curves\cite{dkk99b}. A separate and
important consequence of neutron decay is that it should also affect the internal shock 
gamma-ray light-curves.
In the proton component internal shocks occur at $r_{sh}\sim c t_v \Gpf^2$, where $t_v$ is 
the variability timescale, and $\Gamma_{pf}$ is the asymptotic proton Lorentz factor.
From energy conservation, for $\eta>\eta_\pi$ this is
$\Gamma_{pf}\sim \eta  (1+\xi)[1-(\xi/[1+\xi])(6/7)(\etspi/\eta)^{4/3}]$, 
and taking into account photon drag one can show that an upper limit is $\Gamma_{pf,max} 
\siml 8.3\times 10^2 \E53^{1/4}\tw1^{-1/4}\ro7^{-1/4}$ $(1+8\xi/7)^{-1/4}$.

The $\gamma$-rays start to arrive at an observer time $t\sim t_v \simg 
10^{-3}t_{v-3}$ s, lasting for a time $t_w$ (where $10^{-3}\siml t_w\siml 10^3$ s). 
For the $n$ component, $r_{sh} \sim c t_{min} \Gamma_{nf}^2$, with $\Gamma_{nf}$ from
equation (\ref{eq:Gnf}) and
$t_{min} \sim \min[t_v , t'_n\Gnf^{-1}]$, where $t'_n\sim 10^3$ s is the comoving frame
neutron decay time. Taking $\xi\sim1$ in the estimates below, 
for $20 \siml \eta \siml \etspi\sim 400$ the neutrons decay and shock beyond the 
proton shock for any $t_v \siml 10^3 \eta^{-1}$, at observer times $t_n=[50s, 3s]$,
while for $\eta\simg\etspi\sim 400$ the neutrons decay and shock beyond the proton shock 
for any $t_v \siml 3 (\etspi/\eta)^{1/3}$ s. 
The typical observed duration of the decay, including the blue shift due to the bulk
motion towards the observer, is $t_n\sim 10^3/\Gnf$, where $\Gnf \simeq \eta$ for 
$\eta<\etspi$ and $\Gnf= (3/4)\etspi (\eta/\etspi)^{-1/3}$ for $\eta\geq\etspi$. Thus 
$t_n$ decreases from approximately 50 s to 3 s for $20\siml \eta \siml \etspi \sim 400$, 
and then slowly increases again as $t_n\sim 3 (\eta/\etspi)^{1/3}$ for $\eta\simg\etspi\sim 
400$, with both $\etspi$ and $t_n$ scaling $\propto [(1+\xi)/2]^{-1/4}$. 

The number of neutron decays is $\propto 1- \exp(-t/t_n)$, so the envelope of the 
neutron-related light curve is the mirror image of a ``fred" (fast rise - exponential 
decay), i.e. an ``anti-fred" (or generally, slow rise - fast decay). 
In general, photon emission starts at $t_v$ from the proton-related component, which 
lasts a time $t_w$ with an arbitrary shape envelope, modulated by spikes of minimum 
duration $t_v < t_w$, depending on the chaotic behavior of the central engine producing 
the outflow. The neutron-related component starts at a later time $t_n > t_v$, and has 
an anti-fred shaped envelope modulated by spikes of $t_v$ and a total duration $t_w$.
If $t_w > t_n$, the anti-fred component would be hard to distinguish because of the 
superposition of the ongoing $p$ and $n$ components. However, for short bursts 
with $t_w < t_n \sim 3$ s, the $p$ and $n$ components are separated: first there is
a pulse of duration $t_w$ with a random envelope, followed after a time $\sim t_n$ by
a pulse with an anti-fred envelope of duration $t_n$, 
and characteristic photon energy softer than the previous by $\epsilon_n/\epsilon_p \sim 
t_w/t_n$ (which if small could be below the BATSE band, but may be detectable with the
Swift satellite\cite{swiftpage}). The latter pulse is a signature for neutron decay in 
the burst.

\section{Discussion}

For characteristic parameters, GRB outflows produce 5-10 GeV $\num\barnum$ and 
$\nue\barnue$ from internal inelastic $p,n$ collisions that create pions.  
The $\nu\barnu$ energy output $E_{\nu\barnu}\sim 5\times 10^{51}\E53 (2\xi/[1+\xi])(2/[1+z])
(\etspi/\eta)^{4/3}$ ergs depends on the total energy $E$ of the GRB and on the neutron 
fraction $\xi$ as well as on the dimensionless entropy $\eta$. 
For a km$^3$ detector, approximately 5 to 10 neutrino events above 10 GeV are predicted 
per year, for a neutron/proton ratio $\xi=1$. These events will be coincident with GRB 
electromagnetic flashes in direction and in time (to an accuracy of $\sim 10$ s), which
can enable their separation from the atmospheric neutrino background.
Underground water detectors of the type being planned by BAIKAL \cite{bai97},
NESTOR \cite{nes98}, ANTARES \cite{ant98} and the Antartic detector ICECUBE \cite{hal99}
could potentially detect these relatively low energy neutrino events if a sufficiently
high density of phototubes were used. About 80\% of these neutrinos are $\num$ and $\barnum$
(in approximately equal numbers) and the remainder are $\nue$ and $\barnue$.
These 5-10 Gev $\nu\barnu$ are followed by $\sim$ 120 MeV $\barnue$ from neutron decay, but 
the event rate from neutron-decay neutrinos is very low. The higher energy neutrinos are 
produced for neutron/proton ratios $\xi >0$ when the dimensionless entropy $\eta= 
L/{\dot M}c^2$ exceeds $\etspi \simeq 4\times 10^2 \L52^{1/4}\ro7^{-1/4}$ $(2/[1+\xi])^{1/4}$,
and are accompanied by $\sim 10$ GeV photons which may be detectable in low redshift cases
with GLAST \cite{glast99}. For a typical GRB at redshift $z\sim 1$ the number fluences in 
10 GeV neutrinos are $N(\barnue+\nue)\sim 0.5 N(\barnum+\num)$ $\sim 10^{-4}$ cm$^{-2}$, 
and one order of magnitude less for GeV photons.

In all bursts where $\xi >0$ the lower energy ($\sim 120$ MeV) neutrinos are produced, 
and neutron decay occurs on an observer timescale $t_n\sim 3 \L52^{-1/4}\ro7^{1/4} 
[(1+\xi )/2]^{1/4} (\eta/\etspi)^{1/3}$ s. For outflows of duration $t_w$, these decays 
will be associated with MeV electromagnetic pulses of duration $\min [t_n,t_w]$, 
which are additional to the MeV pulses expected from shocks in the original proton component. 
For short bursts with $t_w \siml 3$ s, the proton electromagnetic pulse appears first and 
is separated from a subsequent neutron electromagnetic pulse, the latter having a slow 
rise-fast decay envelope and a softer spectrum, which may be detectable with the Swift
satellite\cite{swiftpage}.
A systematic study of the time histories of GRB  emission would be useful to search for
evidence of delayed pulses that might be caused by neutron decay.

The detection of 5-10 GeV $\nu\barnu$ in coincidence with GRB photon flashes will not 
be easy, but would provide unique astrophysical information.  Constraints on the neutron 
fraction could provide information about the progenitor stellar system giving rise to GRB. 
For instance, core collapse of massive stars would lead to an outflow from an $Fe$-rich
core with $\xi\sim 2/3-1$, while neutron star mergers would imply $\xi \geq 1$. 
Photodissociation during collapse or merger, as well as $n,p$ decoupling and inelastic 
collisions, would both drive $\xi$ toward unity, although this equalization process is likely 
to remain incomplete. For low $\eta\siml \etspi$, inelastic collisions are not expected 
and the 5-10 GeV $\nu\barnu$ are absent, producing only the harder to detect $\sim 100$ MeV 
$\barnue$ from neutron decay. An initially non-baryonic outflow of, e.g. $e^\pm$ and 
magnetic fields, would acquire a baryonic load by entrainment from the progenitor 
environment, with $\xi \ll 1$ from massive stellar envelopes, but $\xi\simg 1$ for, e.g.,
compact mergers. Thus, lower values of $\xi$, leading to lower ratios of 5-10 GeV 
$\nu\barnu$ and a lower ratio of neutron decay MeV photons to total fluences would be 
expected from massive progenitors than from compact mergers. 

{\it Acknowledgements}\\
Partial support was received by JNB from NSF PHY95-13835 and by PM from 
NASA NAG5-2857, the Guggenheim Foundation and the Institute for Advanced Study.
We acknowledge valuable conversations with G. Fishman, V. Fitch, Vl. Kocharovsky,
P. Kumar, R. Nemiroff, J. Norris, M.J. Rees, P. Vogel and E. Waxman.

\end{document}